\documentclass[twocolumn,showpacs]{revtex4}
\usepackage{graphicx,wrapfig}
\usepackage{dcolumn}
\usepackage{amsmath}
\usepackage[latin1]{inputenc}

\begin{document}

\title{Exploring Valleys of Aging Systems: \mbox{The Spin Glass} Case}
\author{Jesper Dall}
\author{Paolo Sibani}
\affiliation{Fysisk Institut, Syddansk Universitet--5230 Odense M, Denmark}

\date{\today}

\begin{abstract}
\noindent 
We present a statistical method for complex energy landscape exploration
which provides information on the metastable states---or 
valleys---actually explored by an unperturbed
aging process following a quench. Energy fluctuations of
\emph{record} size are identified as the events which move the system 
from one valley to the next. This allows for a semi-analytical
description in terms of log-Poisson statistics, whose main features are
briefly explained. The bulk of the paper is devoted to thorough 
investigations of Ising spin glasses with Gaussian interactions of 
both short and long range, 
a well established paradigm for glassy dynamics.
Simple scaling expressions with universal exponents for 
(a) barrier energies, (b) energy minima, and (c) 
the Hamming distance as a function of the valley index are
found. The distribution of residence time inside valleys entered at 
age $t_w$ is investigated, along with the distribution of 
time at which the global minimum inside a valley is hit. 
Finally, the correlations between the minima of the landscape are 
presented.

The results fit well into the framework of available
knowledge about spin glass aging. At the same time 
they support a novel interpretation of thermal relaxation 
in complex landscapes with multiple metastable states. 
The \emph{marginal stability} of the attractors selected is
emphasized and explained in terms of geometrical properties
of the landscape. 
\end{abstract}


\maketitle

\section{Introduction} \label{introduction}

Physical properties of glassy systems quenched from a high temperature 
slowly change with the time or age $t_w$ elapsed since the quench. 
For any $t_w$ well below the true equilibration time, the dynamics
deceptively appears to be stationary when observed
on time scales shorter than $t_w$. 
Experimental and numerical evidence for the presence of 
a quasi-stationary fluctuations regime for $t \ll t_w$ 
followed by non-equilibrium drift for $t \gg t_w$, 
stems e.g.\ from measurements of conjugate linear response
and autocorrelation functions~\cite{Reim86,Refrigier87c,Andersson92} which 
obey, respectively violate, the fluctuation-dissipation theorem
in the two regimes. In general, the (apparent) age of the system can be 
deduced from the decrease in the rate of change of macroscopic averages. 
This apparent age can however be `reset' to an earlier value 
by applying a perturbation of short duration, as e.g.\ a temperature
pulse, which thus rejuvenates the system. 
 
Non-equilibrium memory effects such as aging and rejuvenation 
were first noticed and studied in a spin glass 
context~\cite{Granberg88,Granberg90,Schultze91,Hoffmann97}, 
but are now observed in a variety of glassy 
systems~\cite{Kityk02,Nicodemi01,Sibani01,Bureau02,Hannemann02}. 
Expanding a previous brief exposition~\cite{sibani02a}, we 
describe and test a statistical approach to landscape explorations
designed to find the generic properties of the
energy landscape which produce these effects. 

Our main point is that the dynamical
events marking the transition between the quasi-equilibrium and the 
off-equilibrium dynamical regimes are triggered by the 
attainment of energy values of record magnitude. 
This immediately allows a description of the non-equilibrium dynamics 
in terms of a log-Poisson process~\cite{Sibani93a,sibanidall03a}, 
i.e.\ a stochastic process which is homogeneous when 
viewed on a logarithmic time scale. 

The sequel is organized as follows: the next section 
introduces the landscape exploration method. Section~\ref{log-Poisson}
briefly explains the relevant properties of 
the log-Poisson statistics used throughout the paper
as a semi-analytical description of non-equilibrium
dynamics. Section~\ref{results} presents results of an extensive 
application of the method to spin glass systems, demonstrating 
the viability of the method and the usefulness of the log-Poisson
description. Section~\ref{Mesoscopic} puts the results in a broader 
perspective, with special reference to coarse-grained mesoscopic 
models of configuration space. In particular, we discuss the 
connection between aging in thermalizing systems and
in dissipative driven system and biological evolution.
Section~\ref{Discussion} is a summary and an outlook. 

As the nature of true equilibrium and the final stages of thermal 
relaxation are only weakly related to the presented data, no theoretical 
consideration is given to these aspects in the paper.

\section{The Method} \label{themethod}
In models of driven dissipative systems with
multiple attractors~\cite{Sibani01,Sibani93a,Coppersmith87},
marginally metastable attractors with an \emph{a priori} 
negligible statistical weight 
are nevertheless those typically selected by the dynamics.
In such systems, this mechanism underlies memory and rejuvenation effects 
analogous to those observed in the thermalization
of e.g.\ spin glasses~\cite{Jonason98} after a quench.
One can speculate that similar mechanisms could generally be present 
in glassy systems with an extensive
number of metastable attractors.
However, the issue of attractor selection is not explicitly considered 
in widely used landscape exploration methods such as the Stillinger-Weber
approach~\cite{Stillinger83,Nemoto88,Becker97,Crisanti02,Mossa02}, 
which study a set of local energy
minima (inherent states) generated by quenches. 
The same applies to exhaustive landscape
exploration techniques~\cite{Sibani93,Sibani98,Schon96} and 
 studies of the real space morphology of 
low-energy excitations by techniques requiring 
quenches~\cite{Andersson96}, genetic algorithms~\cite{Palassini99}
and energy minimization of excitations of 
fixed volume~\cite{Houdayer00,Lamarcq02}.
 
The ability of small external perturbations to 
induce strong rejuvenation and memory effects
in complex dynamics strongly suggests that 
any probe introducing extraneous 
elements in the dynamical evolution might at the same time  
yield a biased picture of the energy landscape.
In other words: are the attractors identified
also those which would be selected by e.g.\ the
unperturbed thermalization process after a deep quench or any
other dynamical evolution of interest?
Analyzing the regions of state space surrounding 
intrinsic states provides valuable information about the
quasi-equilibrium (fluctuation) dynamics
in the energy landscape, but it does not tell the whole story.
The question of how to properly describe
the non-equilibrium process of `selecting'
the metastable states is still open.
This question motivates the present approach 
which is based solely on statistical information 
collected during an \emph{undisturbed aging process}. 

Conventionally, a valley is a 
connected neighborhood of configuration space 
which supports a state of approximate thermal 
equilibrium centered on a local energy minimum. 
During the time a trajectory `resides' in a valley,
the energy and other physical quantities 
fluctuate around a fixed average, and the state of lowest energy
is often revisited; the dynamics has a recurrent character. 
Non-equilibrium events, henceforth `quakes', 
move the system irreversibly from the 
neighborhood of the initial local energy minima of high value
and into progressively deeper valleys. A sequence of 
such events seldom or never 
revisits the same configurations and has a transient character. 

As widely recognized, the lack of time translational 
invariance in aging systems 
stems from the dynamical in-equivalence of the valleys visited.
Consider therefore a landscape with multiple valleys of varying
degrees of metastability, or depth.
On time scales larger than the residence time
of the deepest (i.e.\ most stable) valley seen up 
to the age $t$, all valleys 
shallower than this valley are, by definition, unstable and 
hence irrelevant for the non-equilibrium dynamics. 
The interesting dynamical objects 
are thus the valleys deeper than the deepest valley 
seen up to time $t$. This points to 
energy records as prospective markers
of the non-equilibrium events. 
 
We shall use the term `energy barrier' to denote the
difference between the energy of the current state
and the lowest seen or best-so-far  
energy minimum. The lowest energy minimum value and 
the highest energy barrier observed 
up to time $t$ will identify the valleys
as they successively appear in the landscape. 

Our classification procedure of the undisturbed
dynamics is as follows: We save the minima and barriers of record 
values encountered and the times at which they occur.
We stipulate that a valley is entered at time $t$ if 
the barrier record achieved at time $t$ happens to be 
the last one prior to the attainment 
of a energy minimum record. To leave a valley, a barrier
record must again be followed by a record in the lowest energy.
Whenever several minima records are found between two
barrier records, we only keep the latest, and therefore deepest,
minimum. 

In short, we operationally
identify the valleys encountered by a series of minima records with at
least one barrier event between them. Note that this
selection procedure must be performed retrospectively, since it
is impossible to know `on the fly' whether a new valley
has been entered or not.
We also stress that the barrier records will not necessarily 
coincide with the lowest barrier separating two consecutive energy minima 
records, as the dynamics, due to entropic effects, 
is not likely to follow the path of lowest energy, an
effect noticed by Wevers et al.~\cite{Wevers99} in the 
landscape of metastable ionic compounds. 

The discovery of a new record in low energy is a non-equilibrium event.
However, by no means does it
imply that the internal dynamics in a valley is entirely 
equilibrium-like. Several sub-valleys are typically explored 
before the energy minimum is encountered which 
eventually remains as the lowest state within the valley.
Only then does the dynamics acquire the recurrent, fluctuation-like
nature which is characteristic of quasi-equilibrium.

Resetting the highest barrier to zero 
at an arbitrary point in time produces numerous barrier records 
but no new valleys before a lower energy value is again recorded.
However, resetting both the record energy and barrier 
values to zero may result in a series of new records being registered, 
which describe sub-valleys within the valley originally explored.
By repeating the simulation with the exact same random numbers, 
this procedure allows one to take a closer look at the internal 
structure of a valley if the resetting is done at the time of entry.

The method presented is generally applicable, easily implemented, 
and does not add much to the total runtime of a simulation.
If the energy landscape explored is simple, e.g.\ if it contains one large,
structureless valley or if it is perfectly periodic, 
our scheme only detects a single valley, since degenerate and hence 
dynamically equivalent minima are appropriately lumped together. 
In other words, our method produces simple results when used on
simple systems. In the following sections we show that non-trivial
results are indeed the outcome when complex landscapes are explored.

\section{Log-Poisson statistics} \label{log-Poisson}
In this section we introduce and motivate the log-Poisson statistics
used through the paper as 
an idealized analytical description of non-equilibrium
events---such as the quakes which lead to new valleys.
We mainly focus on the consequences and predictions for 
various quantities of physical interest, leaving the 
empirical justification of the formalism 
to the next section, where the data are also presented.

A log-Poisson description of complex dynamics was first 
introduced in connection with a model 
of Charge Density Waves~\cite{Sibani93a}, and later used to explain 
macro-evolutionary patterns from the fossil record~\cite{Sibani95,Sibani98a}.
It describes~\cite{Sibani99a} the coarse grained
dynamics of a population evolving in a 
NK landscape~\cite{Kauffman87}, and 
there are indications that it might also apply to 
far more realistic models of evolution~\cite{Hall02}. 

The familiar Poisson process with the time argument 
replaced by its logarithm is in short denoted log-Poisson:
\begin{equation}
P_k(t) = \frac{( \alpha \ln t )^k}{k!} 
t^{-\alpha}, \ \ t \geq 1.
\label{simple_logpoisson}
\end{equation} 
As shown in~\cite{Sibani93a,Sibani98a}, the probability 
that $k$ records occur in a sequence of $t$ random 
numbers is given by Eq.~(\ref{simple_logpoisson}), 
independently of the underlying process generating the numbers. 
Log-Poisson statistics implies that the tempo 
at which the events occur falls off as $1/t$. 
Switching to $\log t$ as the independent variable
gives a constant (logarithmic) rate of events,
and restores time homogeneity. Several other interesting 
mathematical properties of log-Poisson processes  
are listed below together with their physical
implications for the spin glass problem at hand. 
 
The probability that $k$ `events' occur between 
$t_w$ and  $t_w+t$ is~\cite{sibanidall03a} 
\begin{equation}
P_k(t_w, t_w+t) = \frac{1}{k!} 
\left[ \alpha \ln \left( \frac{t_w+t}{t_w} \right) \right] ^k 
\left[ \frac{t_w+t}{t_w} \right] ^{-\alpha}.
\label{distribution}
\end{equation} 
Consider a function $c(m,m+k)$ describing the
effect of $k$ events, for example the overlap between the
configurations of lowest energy in valleys $m$ and $m+k$.
To connect with the thermal correlation $C(t_w, t_w +t)$,
we can utilize  a cartoon rendering of the coarse-grained
non-equilibrium dynamics as a one dimensional 
walk on a set of states corresponding to the minima of the valleys,
which  is similar to e.g.\ one of the basic assumptions of Ref.~\cite{rinn00}.
We assume, in our terminology, that the 
dynamical trajectories  of an aging system dwell
at the bottom of the `current' valley for a certain  residence time,
until a quake instantaneously moves  them 
to the bottom of the  next valley. This  coarse grained    picture  
neglects the internal structure of the valleys,  
which is questionable since the time spent searching 
for the bottom  state within a valley
is of the same order of magnitude as the full residence time, see 
e.g.\ Fig.~\ref{rel_time}. 
Nevertheless, the cartoon has the virtue of simplicity, and allows us to 
write 
\begin{equation}
C(t_w, t_w+t) = 
\sum_{m,k} P_m(t_w) P_k(t_w, t_w+t)c(m,m+k).
\label{one-point}
\end{equation} 
which is a function of $t / t_w$ if and only if $c(m,m+k)$ is 
independent of $m$. 
This \emph{pure} or \emph{full} aging form
has very recently been shown to describe aging in real spin glasses,
if cooling rate effects through the critical temperature are 
accounted for~\cite{rodriguez02}. 
The assumption that $c(m,m+k)$ only
depends on a single argument has been checked separately, 
and an additional dependence on the valley index $m$
has been found, which entails a deviation from 
$t / t_w$ scaling~\cite{dall03b} for the numerical models.

For later convenience we finally note that,
if $t_k$ marks the time of the $k$'th event in a log-Poisson process
occurring after $t$, the `log-waiting time' $\log(t_k) - \log(t)$ is 
exponentially distributed, in perfect analogy to
the usual Poisson process. The same exponential distribution also 
describes the series of \emph{independent} 
variables $\log(t_k)-\log(t_{k-1})$. This 
property of the log-waiting time distribution
is very well fulfilled by our data.

\section{Models and Dynamics} \label{Models}
As glassy dynamics is insensitive to many details of the interactions, 
computational convenience is a prime criterion for choosing a 
test model. Spin glass models are comparatively easy to simulate 
and have been investigated
experimentally and numerically for more than twenty years. 
The lack of a comprehensive and coherent picture of their dynamics 
and statics furthermore endows them with considerable intrinsic interest.

This section demonstrates that 
a simple and consistent geometrical picture of the spin glass energy 
landscape is obtained with our `non-invasive' exploration
method. Current spin glass issues are mentioned 
as needed, while a more complete discussion can be found in 
Section~\ref{Mesoscopic}.

We consider $N$ Ising spins, where the energy of a 
spin configuration $\mathbf{s} = \{s_1,\ldots,s_N\}$ is given by
\begin{equation}
E(\mathbf{s}) = - \frac{1}{2}\sum_{i,j} J_{ij} s_i s_j,
\label{SGenergy}
\end{equation}
where the couplings $J_{ij}$ are symmetric, 
independent Gaussian variables of unit variance. 
To test the importance of the topology of the system, we have simulated 
lattices with periodic boundary conditions in both $2d$, $3d$ and $4d$ ($N=L^d$), 
where $J_{ij}$ are non-zero only if $i$ and $j$ are lattice neighbors.
Additionally, we have placed the spins in a $k$-regular random graph, i.e.\ 
each spin interacts with exactly $k$ spins chosen at random.
 
We use single spin flip dynamics coupled with a 
rejectionless algorithm, the Waiting 
Time Method~\cite{Dall01}. The latter
generates a sequence of moves 
equal in probability to the sequence of \emph{accepted} moves 
in the Metropolis algorithm.
Thus, the results of this paper can also be 
obtained with the standard Metropolis algorithm, 
albeit at the price of considerably longer run times. 
The `intrinsic' and size independent 
time variable $t$ used throughout corresponds to the number
of Monte Carlo (lattice) sweeps in the Metropolis 
as well as to the physical time of a real experiment.
We refer to~\cite{Dall01} for a detailed account 
of the Waiting Time Method. 
In all our runs, we conventionally skip the data within the first $10$
time units in order to let the system settle down from 
the random $T = \infty$ initial configuration.

\section{Results} \label{results}
\begin{figure}[t]
\begin{center}
\includegraphics[width=7.5cm]{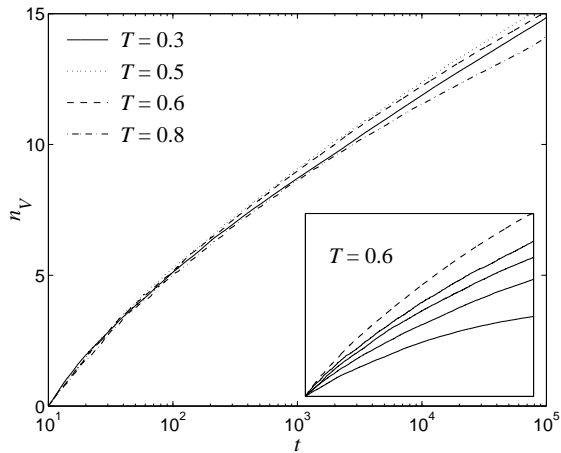}
\vspace{-0.2cm}
\caption
{ \small 
The average number of valleys $\overline{n_V}$ visited
as a function of time in $3d$ Gaussian 
spin glasses with $N=16^3$ spins. For $t>100$ a curvature is barely discernible. 
The insert shows $\overline{n_V}$ for system sizes 
$L = 6, 8, 10$ and $12$ (full lines), as well as $L=16$ 
(dashed line as in the main panel).
}
\label{no_valleys}
\end{center}
\vspace{-0.5cm}
\end{figure} 
Figure \ref{no_valleys} shows the average number of valleys 
$\overline{n_V}(t)$ (which is equal to the number of quakes leading 
from a valley to another) observed in $3d$ at various low temperatures. 
Neglecting the earliest part of the simulations where the memory of the 
initial random spin configuration is still present, 
$\overline{n_V}(t)$ has a logarithmic shape.
As the ratio of the variance $\sigma^2_{n_V}(t)$ to $\overline{n_V}(t)$ 
is constant and close to one (see Fig.~2 of ref.~\cite{sibanidall03a}), 
the statistics of valleys is essentially a 
$\log$-Poisson process, as expected in a record induced 
dynamics~\cite{Sibani93a}. As this holds true for barrier records 
as well (not shown), the ratio of the number of barrier
records to minima records is therefore constant on average.
This non-trivial geometrical feature of the spin glass landscape has not 
previously been noted, and it provides a link to an analytical description 
of the non-equilibrium dynamics as a log-Poisson 
process~\cite{Sibani93a,sibanidall03a}. 

The barely perceptible curvature seen 
in Fig.~\ref{no_valleys} for large $t$
decreases systematically as the system size increases at a fixed temperature, as 
shown in the insert. 
The sub-logarithmic form of $\overline{n_V}(t)$ means that 
the likelihood that a record in low energy
follows a barrier record decreases very slowly, but 
systematically as the system ages. Taking into 
consideration that the curvature seems to vanish 
in the limit of a very large system (see Fig.~2 of ref.~\cite{sibani02a}, 
which shows that the slope of the straight part of $\overline{n_V}(t)$
seems to be proportional to $\log(N)$), 
we surmise that its presence reflects the increasing difficulty in 
finding new low energies as the ground state is approached. 

In summary, the series of quakes moving the system from one
valley to the next can be meaningfully idealized 
into a log-Poisson process discussed in the
previous section, if one disregards the curvature of the data.
How record events are distributed in time 
is insensitive to the properties
of the stochastic process from which the
records are drawn~\cite{Sibani93a}. In our case
Figs.~\ref{no_valleys} and \ref{residence_t}
have a weak temperature dependence, as opposed to 
the strong $T$ dependence of the underlying fluctuations. 
Finally, we note that identical results with respect to 
$\overline{n_V}(t)$ are also found in $2d$, $4d$ and random graphs.

The rest of this section deals with two scaling plots
similar to those presented in ref.~\cite{sibani02a} 
(Figs.~\ref{scaling_B} and \ref{scaling_E}), as well as new results
concerning the scaling of Hamming distances (Fig.~\ref{scaling_H})
between the local minima of contiguous valleys. Finally, 
the statistics of the residence time and the fraction thereof 
spent before hitting the lowest energy state in the valley is 
analyzed, along with the correlations between the minima of the valleys.

\subsection{Scaling of barriers}
\label{barriers}
The data shown in this section are averages over many thousands of 
realizations of the couplings $J_{ij}$ and cover a wide
range of low temperatures and system sizes.
The raw data for each $T$ and
$N$ have very little scatter, and the main sources of error are systematic. 
For example, our finite runtimes of $t=10^6$, combined with 
the very broad distribution of residence times in the valleys, 
bias the average energy of valleys discovered
late in the process, since only the 'faster' trajectories are able to 
explore these. The data are parametrized by the valley index $i$.

In our plots the scaling of the ordinate is 
given in the corresponding labels. The value of the valley index $i$ 
is shifted from one data set to another by up to one unit. 
This corresponds to a multiplicative shift 
of the age, and compensates for 
the arbitrariness of skipping data within the first 10 time units after
the quench, irrespective of temperature and system size. 
 
\begin{figure}[t]
\begin{center}
\includegraphics[width=7.5cm]{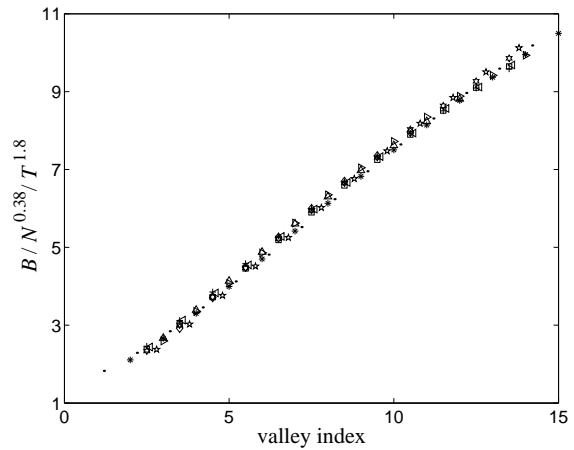}
\vspace{-0.2cm}
\caption{
\small Average energy barrier records $B$ separating 
contiguous valleys in random graph spin glass landscapes 
are scaled with system size $N$ and temperature $T$ as indicated in the 
ordinate label and plotted versus the valley index.
The data points are based on $N=4096$ and $T=0.3,...,0.7$ 
as well as $T=0.5$ and $N=1000,...,16000$.
The same scaling form $B \sim N^{0.38}T^{1.8}$ applies   
equally well to our $2d$, $3d$, and $4d$ data.
}
\label{scaling_B}
\end{center}
\vspace{-0.5cm}
\end{figure} 

Figure~\ref{scaling_B} is a scaling plot of the energy barrier $B_i$ 
which on average must be surmounted in order to leave the $i$'th valley.
The system is a $k$-regular random graph with $k=6$, i.e.\ the same 
number of links per spin as in $3d$. 
$B_i$ is seen to scale in a simple way with the temperature and the 
size of the system:
$B_i(N,T) = B_i(N^aT^b)$, 
where $a = 0.38(2)$ and $b = 1.8(1)$. 
Interestingly, the very same values of $a$ and $b$ are found when scaling 
the barrier heights in $2d$, $3d$ and $4d$ lattices as well 
(see e.g.\ Fig.~3 in ref.~\cite{sibani02a}). 
So, in addition to the non trivial fact that a straightforward two 
parameter fit works very well, the values of the scaling exponents seem  
universal. As the next section will reveal, this universality is also 
found when looking at energy minima and the Hamming distance between them.

The close to linear shape of the scaling plot 
shows that the barrier heights between valleys gradually increase.
Again, we stress that the barriers discussed are those found 
along an unperturbed trajectory, end hence unlikely to be the 
lowest barriers between energy minima in contiguous valleys.

Since the exponent $a$ is positive, the quakes 
do not remain localized to a finite set of spins 
in the macroscopic limit. Some information on the
size and shape of the quakes can be inferred from 
the scaling law for the energy barrier, if we assume that 
quakes correspond to the motion through the system of a generalized 
domain wall. Letting $m$ be the number of
spins typically involved in a quake, we write 
$B \propto m^x$ for some exponent $x>0$. Were the barrier energy 
the outcome of a fluctuation, i.e.\ the sum of $m$ contributions of 
random sign, the behavior would be diffusion-like, with $x=1/2$. However, 
since the barrier crossing process favors 
low-energy states, we expect a lower barrier energy, i.e.\ $x < 1/2$.
As $m \propto N^{0.38/x}$ and $m \leq N$,
we conclude that $x \in [0.38, 0.5]$. In summary, quakes have a fractal 
shape with exponent $0.38/x$, and are more space
filling than a conventional domain wall in $d \leq 4$. 
 
An explanation of the strongly non-Arrhenius $T$ dependence of the 
exponent $b$ involves entropic effects and the connection between the 
quasi-equilibrium dynamics in configuration space and in real space.
The qualitative argument given below 
predicts that the barriers grow linearly with the valley index, and 
that $b = 2$ which is close to the actual value of $b = 1.8$. 
The discrepancy might arise because we neglect that 
domain sizes are distributed quantities.
 
In real space, the relevant objects are connected 
domains~\cite{Andersson96,kisker96} 
of thermally equilibrated spins which slowly grow in a sea of frozen 
spins reaching, on average, a linear length scale $\ell(t)$
on a time scale $t$.
In the progressively longer quiescent periods between the quakes, 
these domains do not interact, and their quasi-equilibrium properties 
are therefore determined by the local density of states ${\cal D}(E)$ 
pertaining to the configurations accessible to the spins constituting 
each domain. ${\cal D}(E)$ has been investigated by 
means of the lid-algorithm~\cite{Sibani93} for a number of different
glassy systems~\cite{Sibani98,Sibani94,Sibani99,Schon98,Schon00}, and 
has consistently been found to have a close to exponential shape: 
${\cal D}(E) \approx \exp(E/T_0(\ell))$. The energy scale $T_0(\ell)$ is 
an upper bound for the temperature at which metastability can 
hold and decreases monotonically with the linear size 
of the system considered~\cite{Sibani98}. While the results quoted 
pertain to systems of fixed size and shape, preliminary investigations 
confirm that the same behavior applies to 
spin domains of size $\ell(t)$ which grow within a large system. 
In this case $T_0(\ell)$ becomes a slowly decreasing function of 
time through the time dependence of $\ell(t)$, and asymptotically 
approaches $T$ from above. We anticipate that the marginal stability 
of the valley visited implies that a domain is typically close to 
its maximal size, i.e.\ $T_0(\ell(t)) \approx T$. 

Returning to $B_i$, we recall that energy barriers delimiting 
valleys are extremal values in a series of ${\cal O}(t)$ independent 
outcomes in a system of age $t$, and that each attempt sees 
energy $E$ with probability 
\begin{equation}
P_{eq}(E) \approx \exp \left( -E \frac{T_0 - T}{T_0 T} \right).
\end{equation}
It follows~\cite{Leadbetter83} that the typical energy barrier scales as
\begin{equation}
 B \propto \frac{T_0 T}{T_0 - T} \log t.
\end{equation}
Since the valley index grows linearly with $\log t$, the linear 
dependence on the valley index seen in Fig.~\ref{scaling_B} is recovered.
Finally, since $T_0\approx T + \epsilon$ for some small $\epsilon$, 
one obtains $b \approx 2$ as anticipated.

\subsection{Scaling of Hamming distances and energies}
\label{HammingEnergy}
\begin{figure}[t]
\begin{center}
\includegraphics[width=7.5cm]{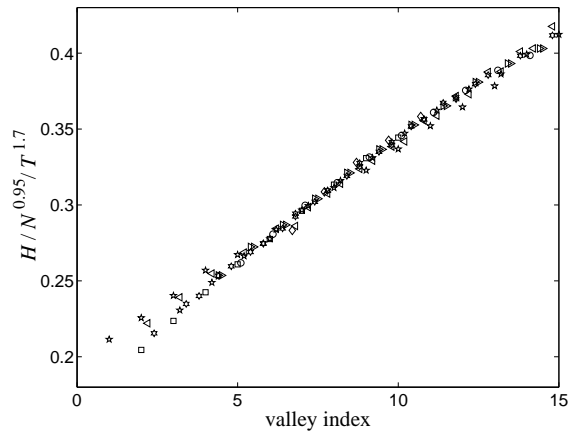}
\vspace{-0.2cm}
\caption{
\small The average Hamming distance $H$ between the lowest energy
configurations of two contiguous valleys in $3d$ spin glasses
is plotted as a function of the valley index.
Combinations of $T=0.5$ and $L=8, 10, 13, 16, 20, 30, 40$ as well as 
$L=16$ and $T=0.4,...,0.8$ are included.
Again, the scaling exponents of $N$ and $T$ are independent of the 
topology of the networks.
} 
\label{scaling_H}
\end{center}
\vspace{-0.5cm}
\end{figure} 
The fact that contiguous valleys contain rather different 
low energy configurations is shown in Fig.~\ref{scaling_H}, which depicts 
the scaling of the Hamming distance $H$ (number of spins which differ 
in their orientation) separating the configurations of lowest 
energy at the bottom of the two valleys in $3d$ lattices. 
Again, a simple scaling relation of the kind 
$H_i(N,T)=H_i(N^aT^b)$ is found in the $d$-dimensional lattices 
simulated as well as in regular graphs. The value of $a=0.95(2)$ 
holds universally, and it indicates that the number of spins involved 
is more or less an extensive quantity. 
The temperature exponent $b$ seems to depend slightly on the topology; 
it has a slightly different value for very low 
temperatures such as $T=0.3$. In short, near-perfect data collapse 
does not seem possible for the whole range of low temperatures simulated. 

\begin{figure}[t]
\begin{center}
\includegraphics[width=7.5cm]{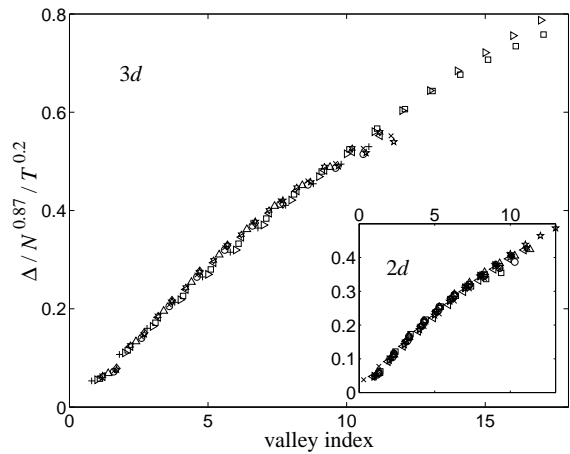}
\vspace{-0.2cm}
\caption{
\small The average
difference $\Delta$ between the states of lowest energy found in the first
and the $i$'th valley is plotted versus $i$. 
The data for $3d$ systems is presented in the main panel, where 
the combinations of $L$ and $T$ shown are as in Fig.~\ref{scaling_H}.
The insert shows that the same scaling exponents work equally well 
in $2d$, where combinations of $T=0.4$ and $L=30,...,100$ as well 
as $L=64$ and $T=0.2,...,0.7$ are plotted. 
}
\label{scaling_E}
\end{center}
\vspace{-0.5cm}
\end{figure}

We must emphasize that data for large values of the valley index must be 
interpreted cautiously when scaling the quantities in 
Figs.~\ref{scaling_B}--\ref{scaling_E}: Since some runs only reach 
a few valleys, averages for larger valley indices become biased. This 
seems particularly important when looking at Hamming distances. Thus, 
we only plot data 
for valleys reached by at least $95\%$ of all runs. By lowering
this threshold to, say, $80\%$, one will clearly see that the extra 
data points lie below the master curve.

The strong temperature dependence of $H$ 
in Fig.~\ref{scaling_H} is remarkable 
considering that the energies of the states involved are virtually
independent of temperature, as implied by Fig.~\ref{scaling_E} 
and as one would expect for actual minima. 
It follows that the observed minima are nearly degenerate, 
which is also expected in spin glasses. The $T$ dependence of $H$ 
is likely due to the fact that at higher $T$
the barriers overcome are higher, and hence the number
$m$ of spins involved in a quake is larger. 
In the simplest scenario where the `downhill' part of the quake 
does not substantially change $m$, one can assume 
$m \propto H \propto N^{0.95}$. The energy of the barrier state, which 
scales as $B \propto N^{0.38}$, must be carried by these $m$ spins. 
With $B \propto m^x \propto N^{0.95 x}$, 
we find $x \approx 0.38/0.95 = 0.4$, in agreement 
with the considerations in the previous section. 

The energy difference $\Delta_i$ between the state of lowest energy 
in the first and the $i$'th valley is plotted in Fig.~\ref{scaling_E} 
for $2d$ and $3d$ lattices. 
$\Delta_i(N,T) = \Delta_i(N^aT^b)$ produces a nice data 
collapse, with $a=0.87(3)$ and $b=0.2(1)$. Again, these exponents 
are found in all topologies simulated, as emphasized by the $2d$ data 
in the insert. The $2d$ data bends slightly more than the $3d$ 
data in the main panel, preventing the ultimate data collapse. This is true 
in general of Figs.~\ref{scaling_B}--\ref{scaling_E}: although the same 
scaling form holds across $2d$, $3d$, $4d$, and $k$-regular random graphs, 
they have slightly different curvatures.

The linear trend in Fig.~\ref{scaling_E} means that 
the lowest energy decreases (almost) logarithmically 
with the age, a feature already implicit in the early investigations by 
Grest et al.~\cite{Grest86}. 
It also implies that the energy difference between neighboring 
valleys remains approximately constant along a trajectory. 
This means that the energy `gain' induced by a quake per 
participating spin is of order $N^{0.87}/N^{0.95} \approx {\cal O}(1)$ 
for the range of system sizes investigated.
In other words, the height of the barrier 
which must be overcome to enter a new valley as well as the overlap 
between them depend strongly on the temperature, while the amount of 
energy gained is nearly independent of $T$. 
Hence, lowering the temperature only slows 
down the process of finding \emph{similar} valleys.
Finally, we mention that the starting point $E_1$ when measuring
$\Delta_i$ is only slightly $T$-dependent. If this was not the
case, we could not claim the similarity of valleys as stated above. 

To sum up, the plots of Figs.~\ref{scaling_B}--\ref{scaling_E} 
tell us that barriers, Hamming distances, and energies of the 
minima of the valleys can all be scaled with respect to size and 
temperature in a simple way. Furthermore, the scaling exponents 
are universal; the same values of the latter are found in 
$d$-dimensional euclidean lattices as well as in regular random graphs.

\subsection{Residence time distribution and superaging}
\label{residence}
The distribution of time spent in `traps' or valleys of the 
energy landscape has, to the best of our knowledge, 
never been measured by others in simulations of spin glass models.
The assumed form of this residence time distribution enters heuristic 
scenarios of spin glass relaxation~\cite{Bouchaud92,Vincent95},
as well as the log-Poisson description of non-equilibrium 
relaxation~\cite{sibanidall03a,dall03b}.

Thus, we consider the probability $R(t \mid t_w)$ that the residence
time $t_r$ in a valley entered at age $t_w$ be less than $t$. As with 
the scaling plots in the previous sections, a clear picture emerges 
which lends support to the usefulness of the method for identifying 
valleys described in Section~\ref{themethod}.

Before we discuss the numerical results in any detail, we present some
theoretical remarks on its expected form for a log-Poisson process. 
Assuming a pure log-Poisson process for $t_w >1 $, 
the `log waiting time' $\ln(t_r+t_w)-\ln(t_w)$ is
exponentially distributed with average $1/\alpha$, where 
$\alpha = d \overline{n_V} / d \ln t$ is the logarithmic slope 
of the curves in Fig.~\ref{no_valleys}. 
For any $x > 0$, the probability of $t_r > t_w(e^x-1)$ is 
therefore $e^{-\alpha x}$. Taking $t = e^x -1 $, the age-scaled 
residence time probability distribution is 
\begin{equation}
R(t \mid t_w) = 1 - \left( \frac{t_w+t}{t_w} \right) ^{-\alpha}. 
\label{R(tw,t)}
\end{equation}
This results also follows directly from Eq.~(\ref{distribution}) by 
noting that $R(t \mid t_w)= 1 - P_0(t_w,t_w+t)$. 
Since $\alpha > 1$~\cite{sibanidall03a}, the distribution has the 
finite average $\langle t_r \rangle = t_w/(\alpha-1)$.
Equation~(\ref{R(tw,t)}) has been shown to fit the empirical residence 
remarkably well~\cite{sibanidall03a}.

\begin{figure}[t]
\begin{center}
\vspace{-0.2cm}
\includegraphics[width=7.5cm]{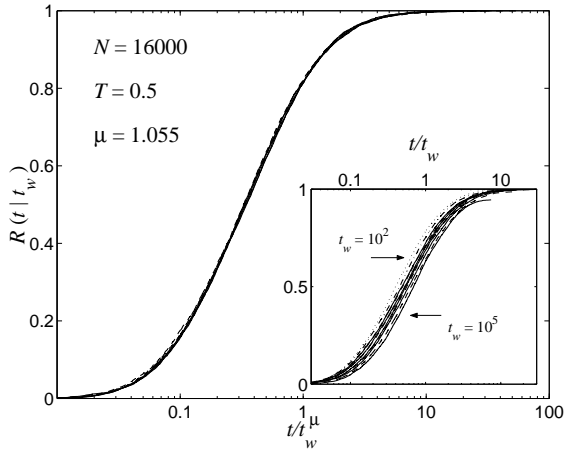}
\caption{ \small 
The main panel shows the probability that the 
residence time $t_r$ is less than $t$ for
valleys entered at $t_w$ for a broad range of $t_w$. 
The abscissa is the scaled variable $t/t_w^\mu$. 
The insert shows that the pure aging $t/t_w$ scaling predicted by 
log-Poisson statistics is a fair but far less satisfactory approximation.
The system is a spin glass on a $k$-regular random graph with $k=6$.
}
\label{residence_t}
\end{center}
\vspace{-0.5cm}
\end{figure}

The $t/t_w$ scaling form is often called pure aging, as opposed to 
super- or sub-aging 
where the scaled variable is $t/t_w^\mu$, with $\mu > 1$ or $\mu <1$,
respectively. Beside Fig.~\ref{residence_t}, superaging of the 
thermal correlation $C(t_w,t_w+t)$ has been observed 
in numerical investigations of spin glasses~\cite{berthier02},
while most experimental data show sub-aging~\cite{vincent97,bouchaud99}. 
Very recent experimental work~\cite{rodriguez02} shows that subaging is a 
consequence of the finiteness of the cooling rate. In the limit of a large 
cooling rate $\mu$ approaches one and pure aging is obtained. 

The empirical data of Fig.~\ref{residence_t} are for 
random graphs of $N = 16000$ spins at $T = 0.5$. As with all the scaling
plots presented so far, the quality of the fit is equally convincing 
in $d$-dimensional lattices.
To study the age dependence of $R(t \mid t_w)$,
the valleys entered are analyzed separately for a broad range of waiting 
time $t_w=10^2,...,10^5$.
As the insert shows, a $t/t_w$ scaling is a fair approximation, 
neglecting the small but systematic drift to the right as $t_w$ grows.

The main panel reveals that an excellent data collapse 
can be obtained with 
\begin{equation}
R(t \mid t_w) = R(t/t_w^\mu), 
\label{all_residence_scaling}
\end{equation}
where $\mu = 1.055$. 
The growth of the residence time with $t_w^\mu$ 
shows that the valleys explored become more stable as 
the system grows older, which concurs with 
the growth law for the barriers given in Fig.~\ref{scaling_B}. 
Physically, this means that, compared to the idealized
log-Poisson case, the residence time in a valley grows
even faster than the age. This deviation from log-time 
homogeneity is likely due to the already mentioned fact that 
it becomes relatively harder to find a new valley
as the ground state is approached~\cite{dall03b}.

Allowing $\mu$ to depend on $T$, the scaling works well for all 
temperatures up to $T = 0.8$. $d\mu/dT>0$ is observed~\cite{dall03b}, 
i.e.\ the superaging effect becomes more pronounced as $T$ increases. 
Noting that simulations slightly below the critical 
temperature, i.e.\ at the high end of the low temperature phase, 
are able to explore lower energy regions in the same span of time, this 
strengthens our hypothesis that deviations from $t/t_w$ scaling
are a reverberation of the finiteness of the ground state energy. 
As such, these deviations can be expected to become less important 
the larger the system is, as they indeed do~\cite{dall03b}.

\subsection{Correlations}
\label{Correlations}
\begin{figure}[t]
\begin{center}
\includegraphics[width=7.5cm]{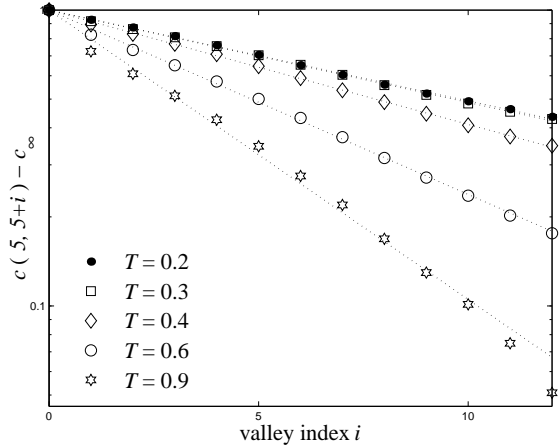}
\vspace{-0.2cm}
\caption{ \small 
The correlation between the valleys $m$ and $m+k$ for $m=5$ in $4d$ 
euclidean lattices of size $N=8^4$. The decay is clearly exponential for 
low $T$. Again, this is observed in all topologies tested.
}
\label{4d_corr}
\end{center}
\vspace{-0.5cm}
\end{figure} 
We have investigated the overlap between configurations $m$ and $m+k$ 
for $m \leq 5$ and $k \leq 12$ and found that the form 
\begin{equation}
c(m,m+k) = c_{\infty}(m) + (1-c_{\infty}(m)) e^{-\gamma(m)k}
\label{corr1}
\end{equation}
fits the data for our range of low temperatures in euclidean lattices as
well as in random regular graphs. Only in $2d$ do we find 
significant deviations, as expected considering that the aging 
is interrupted in this case~\cite{kisker96}. As an example, 
Fig.~\ref{4d_corr} shows $c(5,5+k)$ in $4d$ models. A similar 
plot for $c(1,1+k)$ in $3d$ systems can be found in Fig.~4 of 
Ref.~\cite{sibanidall03a}, where the connection to the experimentally 
available non-equilibrium exponent $\lambda$ is verified. 

If the limiting value for $k \rightarrow \infty$ of $c(m,m+k)$, 
$c_{\infty}(m)$, were independent of $m$, it would coincide with the 
Edwards-Anderson order parameter. Empirically, we find a small 
$m$ dependence, 
for which we have no physical interpretation. The exponent $\gamma$ also 
has a small and almost linear $m$ dependence. 
As Eq.~(\ref{corr1}) remains a heuristic 
approximation of limited value, we have not pursued the size 
and temperature dependence of the parameters involved in 
Eq.~(\ref{corr1}).
Still, we find it interesting that a simple exponential parametrization 
in the number of quakes $k$ accurately describes the data.

\subsection{Hitting time for the minima}
Having argued that our method of partitioning the sampled states
into valleys leads to consistent results and provides useful insight
into the coarse structure of complex energy landscapes, it is 
natural to take a first look at the 
rich \emph{internal structure} of the valleys. 
The existence of such a structure is implied by the
wide distribution of residence times, which requires
a matching distribution of internal energy barriers. Direct 
evidence is presented in this section, where we consider 
the 'hitting time' $t_{hit}$ elapsing from the time of
entry to the time where the state of lowest is encountered. 

Consider $\tau = t_{hit}/t_r$, the fraction of 
the residence time spent 'searching' for the global minimum. 
We expect $\tau$ to be close to zero 
in a structureless valley, where the global minimum is reached 
soon after entry, and close to one in the opposite limit 
of a rugged valley with many internal minima. 
\begin{figure}[t]
\begin{center}
\includegraphics[width=7.5cm]{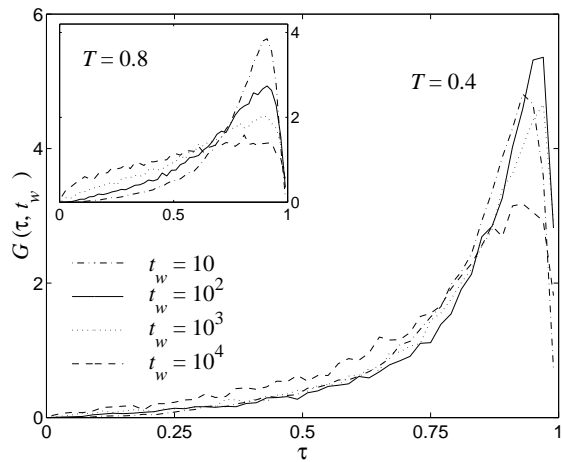}
\vspace{-0.2cm}
\caption{ \small 
The location of the energy minimum within a valley
entered at time $t_w$ in $3d$ spin glasses with $N = 16^3$ spins.
The abscissa is the time relative to the entry and exit times. 
The figure illustrates the highly right-skewed
distribution found for all low $T$, independent of $t_w$.
The insert shows that for higher $T$ the distribution flattens
out as the age $t_w$ of the system grows.
}
 \label{rel_time}
\end{center}
\vspace{-0.5cm}
\end{figure} 
In measuring the distribution of $\tau$, we keep track of the age $t_w$
at the time of entry, by selecting entry times for a range of different $t_w$. 
From the outset, the empirical distribution $G( \tau \mid t_w)$ 
could depend on $t_w$, but, as shown by Fig.~\ref{rel_time}, there is 
practically no dependence, except for $T$ close to the critical temperature. 

The right-skewed form of $G$ for lower $T$ values 
(the picture is the same for any $T \leq 0.6$)
implies that by far the greatest part of $t_r$ is spent
before hitting the lowest minimum: the system explores several 
sub-valleys, each identified by its own local minimum, while
it slides towards states of lower energy. 
Equilibrium-like fluctuations can only occur after 
the state of lowest energy which remains at the bottom 
of the valley is hit. This time interval is small 
compared to the total residence time, but nevertheless grows
steadily with the valley index. 
 
Since the measured shape of $G$ at low $T$ is almost $t_w$ invariant, 
the distributions of both $t_r$ and $t_{hit}$ must 
both scale as $t_w^\mu$. We conclude that the process
of hitting the lowest minimum in a valley at low $T$ is limited 
by internal barrier crossing events, and that these 
barriers themselves grow as the age increases 
and the valleys become larger. The landscape 
geometry is thus invariant under a time dilation,
and hence self-similar, in agreement with the
general properties of a log-Poisson process. 
The temperature independence of the distribution at low $T$ leads to 
the same conclusion: Lowering the temperature means exploring 
smaller valleys, which nevertheless retain 
the same internal structure within a range of low temperatures.

As the temperature gets closer to the critical temperature, a 
waiting time dependence appears: $G(\tau)$ remains peaked around
$\tau \simeq 0.9$, but smaller values of $\tau$ become gradually more 
probable. The corresponding flattening of the distribution of $G(\tau)$ 
for $T=0.8$, a high temperature, is shown in the insert of 
Fig.~\ref{rel_time}. 
Since the number of potentially new valleys diminishes in time because 
of the lack of new low energy records, $G$ must eventually become 
highly \emph{left}-skewed. While only a very slow trend towards 
this situation can be seen in the main panel, the data for the 
higher temperature in the insert seems to be heading in that direction, 
in accordance with equilibration happening much faster at high $T$ than 
low $T$.

We end this section by noting that, like most other quantities presented 
in this paper, similar results for $G( \tau \mid t_w)$ are 
found across all systems simulated.

\section{Coarse grained landscape description} \label{Mesoscopic}
Having argued that log-Poisson statistics is a (slightly)
idealized description which leads to pure aging 
and other dynamical features found in experiments,
we turn to the physical mechanism behind 
the selection of the attractors and, more broadly,
consider the implications of our results for pertinent mesoscopic models
of complex landscapes. We do not attempt to deal with domain growth 
and other real space issues in any detail. 
These aspects, though very important for a 
complete understanding of complex relaxation, 
are too strongly connected to quasi-equilibrium issues
which lie beyond the scope of this paper. 

The applicability of log-Poisson statistics to \emph{barrier} records 
only implies that the trajectory fully decorrelates between successive events,
as can be expected for an activated process in a landscape with many local 
minima. The new and important information 
about the landscape geometry lies in the fact that 
\emph{barrier records are required in order to find new minima records}.
This explains why the log-Poisson statistics 
is equally relevant for barriers as for minima. Secondly, 
and more importantly, it implies that the 
least stable, or \emph{marginally} stable among the
available attractors are those selected. Indeed, if 
at some stage, very deep minima were 
to follow a shallow barrier, subsequent barrier records would 
not produce new states of record-low energy, 
and the log-Poisson description would fail. 
 
The close match between the depth of a valley and the magnitude of the 
barrier record giving access to it implies a
quasi-continuum of available attractors and 
was first noticed in connection with the noisy dynamics of
a driven dissipative system~\cite{Sibani93a}, where the
phenomenon was dubbed noise adaptation.
That complex memory behavior is linked to marginal stability
of metastable attractors has long been known for 
noiseless models of Charge Density Waves~\cite{Coppersmith87,Tang87}. 
The extremely simple automaton 
model of Tang et al.~\cite{Tang87} describes a sheet of elastically
coupled `balls' driven along a sinusoidal potential by 
a pulsed external force.
A recent study of a noisy version of this model~\cite{Sibani01} 
shows that its non-equilibrium dynamics is described by 
log-Poisson statistics and that the age of the system can be reset 
by a change of the elastic coupling constant.

The physical origin of the bias towards shallow attractors 
in thermal glassy dynamics is likely to be entropic, i.e.\ simply 
the fact that shallow attractors vastly outnumber deeper ones, 
in line with the general observation that quenches usually produce
poor minima. To support such bias for a range of low temperatures, 
the density of energy minima must dwarf the Boltzmann factor and 
hence increase at least exponentially with the energy. 
This concurs with the outcome of numerical exhaustive investigations
of the local configuration space structure of different glassy 
systems~\cite{Sibani93,Sibani98,Sibani94,Schon98,Schon00,Klotz98,Schon02}
which were performed with the lid method~\cite{Sibani93,Sibani99}. 
In all cases the local density of states and the local density of minima
are nearly exponential functions of the energy. 

An exponential density of states in connection with activated dynamics implies
a dynamical glass transition. This exact feature was built into the 
tree model of complex relaxation proposed and analyzed by
Grossmann et al.~\cite{Grossmann85} and later studied in more detail 
in~\cite{Hoffmann85,Sibani87}. It is also incorporated 
in the even simpler trap model of Bouchaud~\cite{Bouchaud92} which, 
nonetheless, is very different from tree models in one important respect: 
trees have the lowest connectivity possible for a connected set, 
while each trap of the trap model is connected to all others. 
 
Log-Poisson statistics only applies as long as new and gradually more stable 
valleys remain available to the dynamics. The simplest way of 
modeling inequivalent valleys is through a hierarchy of energy barriers
separating degenerate states, which are organized in either a linear
array~\cite{Joh96} or in a tree graph. Beside the models already mentioned,
the latter approach is followed 
in~\cite{Sibani87,Joh96,Schreckenberg85,Sibani86,Sibani89,Hoffmann90}. 
Many important features of complex relaxation can be reproduced in this
model, but not the fact that in many systems, including spin glasses,
the energy decreases logarithmically with the age. 
This can be achieved by introducing non-degenerate local minima, 
as done in the so-called LS tree~\cite{Hoffmann97,Sibani91}.
The minima have energies which on average decrease
linearly with the size of the barrier overcome, i.e.\ in the 
same overall fashion as Fig.~\ref{scaling_E}. 

In Bouchauds model, trap energies 
are exponentially distributed and hence non-degenerate. 
That deeper minima are gradually explored is a statistical 
consequence of the (assumed) infinite average
of the residence time in a trap. If traps and valleys can
be identified, the results of Section~\ref{residence}
are at variance with this interpretation. The average 
 residence time, which equals $t_w$ for $\alpha =2$,
is, in practice, slightly lower than $t_w$.
This is again reminiscent of the situation encountered in 
 tree models~\cite{Hoffmann97,Sibani87,Schreckenberg85,Sibani86,Sibani89,Hoffmann90,Sibani91}.
 The distribution of barriers in tree models has 
a lower cut-off, unlike 
the fractal description of configuration space of
Dotsenko~\cite{Dotsenko85,Vincent91}, 
 which better captures the growing importance
 of gradually smaller barriers as the temperature decreases. 

A last important issue is the connection between
the energy barrier separating two configurations
and the distance between them. For spin glasses the
relevant metric is the Hamming distance,
which, according to Figs.~\ref{scaling_H} and~\ref{scaling_E}, on
average bears a linear relationship to the energy barrier. A similar result 
was found in numerical work on the SK model~\cite{Sherrington75} 
by~\cite{Nemoto88,Vertechi89}, and by the lid-method
(i.e.\ exact exhaustive enumeration) in~\cite{Sibani94} 
for short range spin glasses. Since the 
latter investigations deal with the small scale structures inside
a `pocket', while the present ones are concerned with the large scale
structures explored by the non-equilibrium dynamics, the agreement
in their outcome is further evidence of a self-similar landscape structure.
Interestingly, the \emph{largest} distance which can be achieved 
for a fixed energy barrier grows exponentially
with the barrier~\cite{Sibani98}. A linear relationship between 
energy and Hamming distance is assumed in tree models of
aging dynamics, see e.g.~\cite{Sibani89}, and also in the 
barrier model~\cite{Joh96}. This is, however, not a crucial
assumption for aging, and other types of functional relationships can
also be utilized~\cite{Hoffmann97} successfully.

\section{Summary and outlook} \label{Discussion}
In this paper we have presented a general `non-invasive' statistical method
for complex energy landscape exploration, especially designed to provide
information on the metastable states actually explored by an unperturbed
aging process following a quench. The method has been thoroughly tested on 
Ising spin glasses, and the results obtained both match and 
extend the established knowledge about spin glasses. In particular, 
most quantities investigated obey simple scaling laws with universal 
scaling exponents.
The view which emerges is that non-equilibrium aging dynamics is steered by 
energy barrier \emph{records}, which are the only 
events capable of opening the route to new valleys.
Since this description has previously been shown to apply to 
driven dissipative models and to evolution 
modeling, a possible unified theory for
non-equilibrium glassy dynamics seems within reach.

A more complete picture can be obtained by looking more closely 
into the fluctuation dynamics which may differ across different systems. 
For spin glasses we have argued that 
real space domains of (pseudo) thermalized spins
relax independently as long as the system as a whole remains in 
the same valley. The present method opens the possibility of identifying the
quasi-equilibrium clusters as defined by the dynamics itself:
These clusters are separated by a backbone of spins
whose orientation remains fixed within each valley and changes slowly
from one valley to the next, as seen in Fig.~\ref{scaling_H}. 
After the lowest energy state in the
valley has been hit and before the next valley is entered, there
is no drift towards the global minimum. Hence, 
the spins fall into two categories only: those which are frozen, and
those which fluctuate in a quasi-equilibrium fashion. The 
quasi-equilibrium clusters can thus be extracted and their statistical
properties, such as e.g.\ the density of states, can be studied for each
cluster separately.

\vspace{0.5cm}
\noindent {\bf Acknowledgments}:
This project has been supported by Statens Na\-tur\-viden\-skabe\-lige 
Forsk\-nings\-r\aa d through a block grant 
and by the Danish Center for Super Computing with computer time
on the Horseshoe Linux Cluster. We are grateful to J. Christian
Sch\"{o}n for many discussions, and, in particular, for pointing 
out a flawed mathematical argument in an early version of this work. 
\bibliographystyle{abbrv}

\bibliography{thesis,SD-meld,sg}
\end{document}